\documentclass[12pt, a4paper]{article}
\usepackage[T2A]{fontenc}
\usepackage[utf8]{inputenc}
\usepackage[english]{babel}
\usepackage{amssymb,graphicx}
\usepackage{amsmath,amsfonts}
\usepackage{amsthm}
\usepackage{fullpage}
\usepackage{hyperref}

 \newcommand{\sgn}{\mathop{\mathrm{sgn}}\nolimits}

\title{Spectral dualities in XXZ spin chains and five dimensional
  gauge theories}
\author{A.~Mironov$^{a,b}$\thanks{mironov@itep.ru, mironov@lpi.ru} ,
  A.~Morozov$^a$\thanks{morozov@itep.ru} ,
  B.~Runov$^{a,d}$\thanks{runovba@gmail.com} ,
  Y.~Zenkevich$^{a,c}$\thanks{yegor.zenkevich@gmail.com,
    zenkevich@ms2.inr.ac.ru} ,
  A.~Zotov$^{a,d,e}$\thanks{zotov@itep.ru, zotov@mi.ras.ru}\\
  {\small $^a$\textit{ITEP, Moscow, Russia}}\\
  {\small $^b$\textit{Theory Department, Lebedev Physics Institute,
      Moscow,
      Russia}}\\
  {\small $^c$\textit{Institute for Nuclear Research of the Russian
      Academy of
      Sciences, Moscow, Russia}}\\
  {\small $^d$\textit{MIPT, Dolgoprudniy, Moscow, Russia}}\\
{\small $^e$\textit{Steklov Mathematical Institute of the Russian
    Academy of Sciences,  Moscow,  Russia}}} \date{}

\begin{document}
\maketitle
\vspace{-55ex}
\begin{flushright}
  FIAN/TD-12/13\\
  ITEP-TH-24/13
\end{flushright}
\vspace{40ex}

\begin{abstract}
  Motivated by recent progress in the study of supersymmetric
  gauge theories we propose a very compact formulation of spectral
  duality between XXZ spin chains. The action of the quantum duality
  is given by the Fourier transform in the spectral parameter. We
  investigate the duality in various limits and, in particular, prove
  it for $q\to 1$, i.e.\ when it reduces to the XXX/Gaudin duality. We
  also show that the universal difference operators are given by the
  normal ordering of the classical spectral curves.
\end{abstract}
Integrable systems provide a key to understanding $\mathcal{N}=2$
supersymmetric gauge theories. The Seiberg--Witten theory~\cite{SW} is
naturally formulated in terms of integrable systems~\cite{SWintsys},
and $\epsilon$-deformation of the former corresponds to quantization
of the latter. Another view on this connection is provided by the AGT
correspondence~\cite{AGT}, in which a (quantization of) Hitchin
integrable system arises in a natural way.

Interestingly, it turns out that for a large class of gauge theories
there are \emph{two different} integrable systems, associated with
each of them~\cite{MMRZZ}. The equivalence between the two systems
is given by the \emph{spectral duality}. This duality was proposed
in~\cite{Harnad1,Harnad2,W1} and further explored in different ways
in ~\cite{MTV01, MTVCapelli, MTVbispec} and  \cite{MMRZZ}. On the
classical level the duality exchanges the two coordinates in the
equation for the spectral curve of the system. The classical
Hamiltonians of the two systems, therefore, coincide. On the quantum
level the situation is more subtle: the duality comes in several
variants. In the weakest version, one claims that some subset of all
quantum Hamiltonians is shared by the two systems~\cite{MTV01}. In a
stronger version~\cite{MTVCapelli}, the whole Bethe subalgebras,
i.e.\ \emph{all} the Hamiltonians, of the two systems are
identified. In the third version~\cite{MTVbispec} one identifies the
orbits of solutions to the Bethe equations.

Here we will consider the second version of the duality and conjecture
the correspondence between the Bethe subalgebras for XXZ spin
chains. Let us note that the eigenvalues of the universal difference
operators provide one with the Baxter TQ equation, which determines
the Bethe equations for the system. Therefore, our approach gives,
albeit indirectly, the duality between the solutions to the Bethe
equations. For the XXX/Gaudin duality this correspondence should
reproduce the results of~\cite{MTVbispec}.

Our approach to quantum duality consists of two steps. First we prove
the classical duality, i.e.\ the equality of the spectral curves of
the two systems and the isomorphism of the Poisson structures. We then
prove that the universal difference operators, representing the
quantization of the spectral curves coincide with the normal ordered
classical expressions (this is actually a statement within the theory
of a particular system and by itself has nothing to do with duality).
It follows that the equality at the classical level is immediately
lifted to the quantum one. To support our claim we prove the duality
in the four dimensional limit, when it reduces to the duality between
XXX spin chain and the trigonometric Gaudin model. In our previous
work~\cite{MMRZZ} we have proven the duality between the XXX spin
chain and the \emph{reduced} Gaudin model. To resolve this seeming
discrepancy we now show the equivalence of trigonometric and reduced
Gaudin systems. For the sake of brevity we present here only the main
ideas of the proofs; more detailed account is left for a future
publication~\cite{MMRZZnew}.

The fact that the ordinary Fourier transform provides an \emph{exact}
bispectral duality between the spin chains is surprisingly close to
the recent claim that it also describes the modular transformation of
conformal blocks in all orders of their perturbative genus
expansion~\cite{GMMNemkov}. Thus, our new result adds to the hope that
quantum dualities can turn out to be much simpler than they seem ---
in appropriate variables they are reduced to Fourier
transform. Moreover, these appropriate variables are also the natural
ones, i.e.\ they are distinguished from the point of view of each of
the dual systems.

The XXZ spin chains correspond to five dimensional quiver gauge
theories and the spectral duality gives rise to the $SU(N)^{M-1}
\leftrightarrow SU(M)^{N-1}$ duality introduced in~\cite{Bao}. Also,
very recently the third version of the duality (the correspondence
between Bethe equations) was also discussed in~\cite{Gaiotto}
employing the ideas of string dualities. We will comment on the
string/M-theory motivation for spectral duality elsewhere. Different
aspects and applications of the spectral duality can be found in
\cite{w2,w3,w4,w5,w6,w7}.

\paragraph{Introducing the XXZ chain.}
\label{sec:introduction-1}

We consider closed XXZ spin chain with $N$ sites and spins in
symmetric representations of $U_q (\mathfrak{gl}_K)$. These
representations will be realised as subspaces of the Fock space
generated by the $q$-Bose creation operators. The antisymmetric case
is completely analogous except for a few signs. In principle,
arbitrary representations can be obtained by the fusion procedure. The
monodromy operator, or transfer matrix, of the chain is given by the
product of the Lax operators residing on the sites of the chain and a
diagonal twist matrix $Q = \mathrm{diag} (Q_1, \ldots, Q_K)$:
\begin{equation}
  \mathbf{T}(v) = Q\, \mathbf{L}^N (v/v_N) \cdots \mathbf{L}^1(v/v_1), \notag
\end{equation}
where
\begin{equation}
  \mathbf{L}^i(v/v_i) = \sum_{a,b} E_{ab} \otimes \left( \delta_{ab}
    q^{\mathbf{N}_a^i + 1}  + \frac{( \delta_{b>a} v + \delta_{a \geq
        b}v_i) (q - q^{-1}) }{ v - v_i} \mathbf{B}_a^i \mathbf{A}_b^i
  \right), \; \;
  \begin{array}{c}
    a,b=1,\ldots,K,\\
    i=1,\ldots ,N,
  \end{array}
  \notag
\end{equation}
and $\mathbf{A}_a^i$, $\mathbf{B}_a^i$ and $\mathbf{N}_a^i$ form the
$q$-Bose algebra,
\begin{equation}
  \begin{array}{rcl}
  \mathbf{A}_a^i \mathbf{A}_b^j &=& q^{\delta^{ij} \sgn(a-b)} \mathbf{A}_b^j
  \mathbf{A}_a^i\,, \\
  \mathbf{B}_a^i \mathbf{B}_b^j &=& q^{\delta^{ij} \sgn(a-b)} \mathbf{B}_b^j
  \mathbf{B}_a^i\,,\\
  \mathbf{A}_a^i \mathbf{B}_b^j &=& q^{\delta^{ij}(\delta_{ab} - \sgn(a-b))}
  \mathbf{B}_b^j \mathbf{A}_a^i - \delta^{ij} \delta_{ab} q^{\mathbf{N}_a+1}\,,
  \end{array}
  \qquad
  \begin{array}{rcl}
    [ \mathbf{N}_a^i, \mathbf{A}_b^j] &=& \delta^{ij} \delta_{ab} \mathbf{A}_b^j\,,\\
    {}[ \mathbf{N}_a^i, \mathbf{B}_b^j] &=& -\delta^{ij} \delta_{ab} \mathbf{B}_b^j\,.
  \end{array}
  \label{eq:1}
\end{equation}
The quantum commuting Hamiltonians $\mathcal{H}_m(v)$ are compactly
collected in the \emph{universal difference operator}~\cite{MTVhigher,
  Hopkins, AlgManin}:
\begin{equation}
  \hat{D}_{N,K}(v, q^{2v\partial_v}) = \sum_{m=0}^K (-1)^m \mathcal{H}_m(v)
    q^{2mv \partial_v} = \sum_{m=0}^K \sum_{A = \{ 1 \leq a_1 < \ldots < a_m
      \leq K\}} \det_{\mathrm{col}} {}_q \left(
      \mathbf{T}_{AA} (v) q^{2v \partial_v} \right),
    \label{eq:2}
\end{equation}
where $\mathbf{T}_{AA}$ is $m \times m$ submatrix of $\mathbf{T}$ and
the \emph{column $q$-determinant} of a submatrix is defined as
\begin{equation}
  \det_{\mathrm{col}} {}_q \mathbf{M}_{AA} = \sum_{\sigma \in
    \mathfrak{S}_m} (-q)^{\mathrm{inv}(\sigma)} \mathbf{M}_{a_{\sigma (1)}
    a_1} \cdots \mathbf{M}_{a_{\sigma(m)} a_m} \,.\label{eq:3}
\end{equation}
The inversion number of a permutation is given by
$\mathrm{inv}(\sigma) = \sum_{a<b} \delta_{\sigma(a)>\sigma(b)}$.

\paragraph{The main conjecture.}
\label{sec:main-conjecture}
We propose the equality between the universal difference operators of
the $N$-site $\mathfrak{gl}_K$ and $K$-site $\mathfrak{gl}_N$ XXZ spin
chains. More explicitly
\begin{equation}
  \label{eq:4}
  \boxed{\prod_{i=1}^N (v - v_i) \hat{D}_{N,K} (v, q^{2v\partial_{v}}) = \prod_{a=1}^K
    \left( 1 - w_a q^{2v \partial_v} \right) \hat{D}_{K,N} (q^{-2v\partial_v},v)}
\end{equation}
In the right hand side $w_a$ are the inhomogeneities of the dual spin
chain.

To make the statement of Eq.~\eqref{eq:4} precise one should also
specify an isomorphism between the operator algebras of the two
systems. This is done as follows. We introduce the ``twisted''
$q$-Bose generators $\widetilde{\mathbf{A}}_a^i$,
$\widetilde{\mathbf{B}}_a^i$:
\begin{equation}
  \begin{array}{rcl}
      \widetilde{\mathbf{A}}_a^i &=& \mathbf{A}_a^i \prod_{j<i}
  q^{\mathbf{N}_a^j}\,,\\
  \widetilde{\mathbf{B}}_a^i &=& \prod_{j \leq i} q^{-\mathbf{N}_a^j}
  \mathbf{B}_a^j\,.
  \end{array}
\end{equation}
The twisted generators satisfy the ``twisted'' $q$-Bose algebra which
is manifestly symmetric under the exchange of the site and algebra
indices $i,j \leftrightarrow a,b$,
\begin{equation}
  \begin{array}{c}
      \widetilde{\mathbf{A}}_{a}^{i} \widetilde{\mathbf{A}}_{b}^{j} =
  q^{\delta_{ij} \sgn(a-b) + \delta_{ab} \sgn(i-j) }
  \widetilde{\mathbf{A}}_{b}^{j}
  \widetilde{\mathbf{A}}_{a}^{i}\,, \\
  \widetilde{\mathbf{B}}_{a}^{i} \widetilde{\mathbf{B}}_{b}^{j} =
  q^{\delta_{ij} \sgn(a-b) +\delta_{ab} \sgn(i-j) }
  \widetilde{\mathbf{B}}_{b}^{j}
  \widetilde{\mathbf{B}}_{a}^{i}\,,\\
  \widetilde{\mathbf{A}}_{a}^{i} \widetilde{\mathbf{B}}_{b}^{j} =
  q^{2\delta_{ij} \delta_{ab} - \delta_{ij} \sgn(a-b) - \delta_{ab}
    \sgn(i-j)} \widetilde{\mathbf{B}}_{b}^{j}
  \widetilde{\mathbf{A}}_{a}^{i} - \delta_{ab} \delta_{ij} q^2\,.
  \end{array}
\label{eq:5}
\end{equation}
We therefore conclude that the action of the duality on the twisted
generators is trivial ($\mathrm{D}$ denotes the dual system)
\begin{equation}
\boxed{
  \begin{array}{c}
      \widetilde{\mathbf{A}}_{a}^{i} =
  (\widetilde{\mathbf{A}}_{a}^{i})^{\mathrm{D}},\\
  \widetilde{\mathbf{B}}_{a}^{i} =
  (\widetilde{\mathbf{B}}_{a}^{i})^{\mathrm{D}}.
  \end{array}
}
  \label{eq:6}
\end{equation}
Having stated our main conjecture we proceed to proving it in several
special cases.

\paragraph{The classical limit.}
\label{sec:classical-limit}
In the classical limit quantum operators $\widetilde{\mathbf{A}}$,
$\widetilde{\mathbf{B}}$ turn into functions $\widetilde{A}$,
$\widetilde{B}$\footnote{One should scale the operators appropriately,
  namely $(q - q^{-1}) \widetilde{\mathbf{A}} \widetilde{\mathbf{B}}
  \to \widetilde{A} \widetilde{B}$.}. The universal difference
operator $\hat{D}_{N,K}(q^{2v\partial_v}, v)$ reduces to the spectral
curve $\Gamma_{N,K}(w,v)$ of the system, i.e.\ the characteristic
polynomial of the monodromy matrix. The Seiberg--Witten differential
can be obtained from the classical asymptotics of the kernel of the
universal difference operator:
\begin{equation}
  \label{eq:7}
  \hat{D}_{N, K} (v, q^{2v\partial_v}) \, Q(v) = 0\,, \qquad Q(v)
  \to \exp \left[ \int^v
    \lambda_{\mathrm{SW}} + \ldots \right].
\end{equation}
More explicitly, the differential is given by $\lambda_{\mathrm{SW}} =
\ln w\, d \ln v$.

For classical systems the duality reduces to the equality between two
determinants which is proven using the elementary identity: $ \det W
\, \det (X - Y W^{-1} Z) = \det \left(
  \begin{smallmatrix}
    X&Y\\
    Z&W
  \end{smallmatrix}
\right) = \det X\, \det (W - Z X^{-1} Y)$. In our case the blocks $X$,
$Y$, $Z$ and $W$ are respectively $K \times K$, $K \times N$, $N\times
K$ and $N \times N$ matrices which are written as follows
\begin{align}
  X_{ab} &= \delta_{ab}w - \delta_{a>b} w_a \sum_{ a_2, \ldots ,a_{N} } (\delta_{aa_N} + \delta_{a_N > a}
  \widetilde{B}_a^N \widetilde{A}_{a_N}^N) \cdots (\delta_{a_2 b} + \delta_{b >
    a_2} \widetilde{B}_{a_2}^1 \widetilde{A}_b^1), \notag\\
  W_{ij} &= \delta_{ij} (v - v_i) - \delta_{i>j} \sum_{a_i,\ldots, a_j}
  \widetilde{A}_{a_i}^i (\delta_{a_i a_{i-1}} + \delta_{a_{i-1} > a_i}
  \widetilde{B}_{a_i}^{i-1} \widetilde{A}_{a_{i-1}}^{i-1} ) \cdots \notag\\
  &\phantom{=} \cdots
  (\delta_{a_{j+1} a_j} + \delta_{a_j > a_{j+1}}
  \widetilde{B}_{a_{j+1}}^{j+1} \widetilde{A}_{a_j}^{j+1} )
  \widetilde{B}_{a_j}^j \,, \notag\\
  Y_{ai} &= w_a \sum_{ a_i, \ldots ,a_{N} } (\delta_{aa_N} + \delta_{a_N > a}
  \widetilde{B}_a^N \widetilde{A}_{a_N}^N) \cdots (\delta_{a_{i+1} a_i} + \delta_{a_i >
    a_{i+1}} \widetilde{B}_{a_{i+1}}^{i+1} \widetilde{A}_{a_i}^{i+1})
  \widetilde{B}_{a_i}^i\,, \notag\\
  Z_{ia} &= \sum_{a_j , \ldots, a_1} \widetilde{A}_{a_j}^j (\delta_{a_j a_{j-1}} + \delta_{a_{j-1} >
    a_j} \widetilde{B}_{a_j}^j \widetilde{A}_{a_{j-1}}^{j-1}) \ldots (\delta_{a_2 a_1} + \delta_{a_1 >
    a_2} \widetilde{B}_{a_2}^2 \widetilde{A}_{a_1}^1). \notag
\end{align}
We thus prove the classical XXZ/XXZ duality. In the next sections we
show that the \emph{quantum} universal difference operators of the
spin chains is given by the normal ordering of the \emph{classical}
expressions, at least in the limit $q \to 1$. Therefore, the classical
result just proven actually implies the quantum spectral duality for
$q \to 1$, i.e.\ for XXX/Gaudin.

\paragraph{XXX/Gaudin duality.}
\label{sec:xxxtr-gaud-dual}

When $q \to 1$ the spectral duality between the XXZ spin chains
reduces to the duality between $N$-site $\mathfrak{gl}_K$ XXX spin
chain and $\mathfrak{gl}_N$ trigonometric Gaudin on a cylinder with
$K-2$ marked points\footnote{The cylinder can be turned into a sphere
  at the expense of adding two more special marked points (``irregular
  singularities'').}. To obtain the XXX chain on the left side of the
duality~\eqref{eq:4} one assumes $v = q^{2x}$, $v_i = q^{2x_i}$ and
keeps $x$, $x_i$ fixed as $q \to 1$. The twisted $q$-Bose operators
$\widetilde{\mathbf{A}}_a^i$, $\widetilde{\mathbf{B}}_a^i$ turn into
the Bose ones $\mathbf{a}_a^i$, $\mathbf{b}_a^i$:
\begin{equation}
  \label{eq:8}
  \begin{array}{c}
    {}[\mathbf{a}_a^i, \mathbf{a}_b^j] = [\mathbf{b}_a^i, \mathbf{b}_b^j] =0\,,\\
    {}[\mathbf{a}_a^i , \mathbf{b}_b^j] = - \delta_{ab} \delta^{ij}.
  \end{array}
\end{equation}
The universal difference operator for the XXX spin chain can be
compactly written as a column determinant\footnote{The definition of
  column determinant is obtained from the definition~\eqref{eq:3} by
  setting $q=1$.}
\begin{equation}
  \label{eq:9}
  \hat{D}_{N.K}(v, q^{2v\partial_{v}}) \to \det_{\mathrm{col}} \left(\mathbf{1} -
    \mathbf{T}_{\mathrm{XXX}}(x) e^{\partial_x} \right).
\end{equation}

On the right side of the duality~\eqref{eq:4} the situation is more
subtle. Unlike the spectral parameter $v$, the $q$-difference operator
$q^{-2v\partial_v}$ does not tend to identity. Thus in the limit $q
\to 1$ the universal difference operator $\hat{D}_{K,N}$ does not
reduce to a column determinant. Instead it gives rise to a complicated
expression, which to our knowledge has not been previously
written. One obtains the following explicit form of the universal
difference operator for the trigonometric Gaudin model:
\begin{equation}
  \label{eq:10}
  \hat{D}_{K,N} (q^{-2v\partial_v},v) \to
  \widetilde{\det_{\mathrm{col}}} \left( x - \mathbf{L}_{\mathrm{tG}}(e^{-\partial_x}) \right),
\end{equation}
where $\mathbf{L}_{\mathrm{tG}}$ is the Lax operator for the
trigonometric Gaudin model and
\begin{multline}
  \label{eq:11}
   \widetilde{\det_{\mathrm{col}}}\, \mathbf{M} =  \sum_{m=0}^{N-2} (-1)^m
  \sum_{J=\{1 \leq i_1 < \ldots < i_m \leq N\}} \; \sum_{\sigma \in \mathfrak{S}_N}
  (-1)^{\mathrm{inv}(\sigma)} \prod_{\alpha =1}^{m} \left(
    \sum_{j=1}^{i_{\alpha} - 1} \delta_{\sigma(j) > i_{\alpha}} \right) \cdot\\
  \cdot \mathbf{M}_{\sigma(1),1} \cdots
  \mathbf{M}_{\sigma(i_1-1),i_1-1} \delta_{\sigma(i_1) i_1}
  \mathbf{M}_{\sigma(i_1+1),i_1+1}
  \cdots\\
  \cdots \mathbf{M}_{\sigma(i_2-1),i_2-1}\delta_{\sigma(i_2) i_2}
  \mathbf{M}_{\sigma(i_2+1),i_2+1}
  \cdots \mathbf{M}_{\sigma(K),K}\,.
\end{multline}
The term with $m = 0$ gives the column determinant, other terms
represent quantum corrections. Notice that for $N=1,2$ the corrections
do not arise. In a moment we will prove that, rather remarkably, all
these corrections exactly cancel with the normal ordering corrections
from the column determinant.

We can now write the XXX/Gaudin duality as follows:
\begin{equation}
  \label{eq:12}
  \boxed{\prod_{i=1}^N (x - x_{i})\det_{\mathrm{col}} \left( \mathbf{1} - \mathbf{T}_{\mathrm{XXX}}(x)
      e^{-\partial_x} \right) = \prod_{a=1}^K \left( 1 - Q_a e^{\partial_x} \right) \widetilde{\det_{\mathrm{col}}} \left( x
      - \mathbf{L}_{\mathrm{tG}} (e^{-\partial_x})\right)}
\end{equation}
In the next sections we employ the normal ordering relations and
classical calculations to prove this duality.

\paragraph{Normal ordering the universal difference operators.}
\label{sec:norm-order-univ}
Let us introduce the notion of normal ordering which will be
instrumental in our proof of the spectral duality:
\begin{equation}
  \label{eq:13}
  \mathopen: F(\mathbf{b}, \mathbf{a},
  x, \partial_x) \mathclose: = \left\{ \parbox{9.9em}{all $\mathbf{b}$
      and $x$
      to the left, all $\mathbf{a}$ and $\partial_x$ to the right}
  \right\}.
\end{equation}
It turns out that the universal difference operator for the XXX spin
chain is equal to its own normal ordering,
\begin{equation}
  \label{eq:14}
  \boxed{\det_{\mathrm{col}} \left(\mathbf{1} -
    \mathbf{T}_{\mathrm{XXX}}(x) e^{\partial_x} \right) = \mathopen: \det_{\mathrm{col}} \left(\mathbf{1} -
    \mathbf{T}_{\mathrm{XXX}}(x) e^{\partial_x}\right) \mathclose:}
\end{equation}
The proof runs along the same lines as that of~\cite{MTVCapelli}. One
notices that $T(x) e^{\partial_x}$ is already normal ordered. The
induction proceeds as follows: performing the necessary contractions
one shows that for any $i$
\begin{multline}
  \label{eq:15}
  \sum_{\sigma \in \mathfrak{S}_K} \left( \mathbf{1} - \mathbf{T}(x)
    e^{\partial_x} \right)_{\sigma(1),1} \cdots \; \mathopen: \left( \mathbf{1} - \mathbf{T}(x)
    e^{\partial_x} \right)_{\sigma(i),i} \cdots \left( \mathbf{1} - \mathbf{T}(x)
    e^{\partial_x} \right)_{\sigma(K),K} \mathclose: =\\
  =\sum_{\sigma \in \mathfrak{S}_K} \left( \mathbf{1} - \mathbf{T}(x)
    e^{\partial_x} \right)_{\sigma(1),1} \cdots \; \mathopen: \left( \mathbf{1} - \mathbf{T}(x)
    e^{\partial_x} \right)_{\sigma(i-1),i-1} \cdots \left( \mathbf{1} - \mathbf{T}(x)
    e^{\partial_x} \right)_{\sigma(K),K} \mathclose:
\end{multline}
Going from right to left in this way one normal orders the whole
expression.

Surprisingly, a similar proof works for the trigonometric Gaudin
universal difference operator. In this case, however, on each step one
obtains corrections, which are exactly canceled by the extra terms in
the definition of $\widetilde{\det}$. The induction now goes from left
to right:
\begin{multline}
  \label{eq:16}
  \sum_{\sigma \in \mathfrak{S}_N} \mathopen: \left( x -
    \mathbf{L}(e^{-\partial_x}) \right)_{\sigma(1),1}
  \cdots \left( x -
    \mathbf{L}(e^{-\partial_x}) \right)_{\sigma(i),i}
  \mathclose: \; \cdots \left( x -
    \mathbf{L}(e^{-\partial_x}) \right)_{\sigma(N),N}=\\
  = \sum_{\sigma \in \mathfrak{S}_N} \mathopen: \left( x -
    \mathbf{L}(e^{-\partial_x}) \right)_{\sigma(1),1}
  \cdots \left( x -
    \mathbf{L}(e^{-\partial_x}) \right)_{\sigma(i+1),i+1}
  \mathclose: \; \cdots \left( x -
    \mathbf{L}(e^{-\partial_x}) \right)_{\sigma(N),N} +\\
  + \sum_{\sigma \in \mathfrak{S}_N} \left( \sum_{j=1}^i \delta_{\sigma(j)>i} \right) \mathopen: \left( x -
    \mathbf{L}(e^{-\partial_x}) \right)_{\sigma(1),1}
  \cdots \left( x -
    \mathbf{L}(e^{-\partial_x}) \right)_{\sigma(i),i}
  \mathclose:\\
  \cdot \delta_{\sigma(i+1),i+1} \cdots \left( x -
    \mathbf{L}(e^{-\partial_x}) \right)_{\sigma(N),N}\,.
\end{multline}
Eventually, one arrives at the following relation
\begin{equation}
  \label{eq:17}
  \boxed{ \prod_{a=1}^K (1 - Q_a e^{\partial_x})\;
    \widetilde{\det_{\mathrm{col}}} \left( x -
      \mathbf{L}_{\mathrm{tG}}(e^{-\partial_x}) \right)  = \mathopen: \prod_{a=1}^K (1 - Q_a e^{\partial_x})
    \det_{\mathrm{col}} \left( x -
      \mathbf{L}_{\mathrm{tG}} (e^{-\partial_x})\right) \mathclose:}
\end{equation}
Notice the simple (without tilde) column determinant on the right hand
side.

What we have shown for the XXX chain and the trigonometric Gaudin
model is essentially that the \emph{quantum} universal difference
operator is given by the normal ordered \emph{classical}
determinant. One can hope that an analogous statement holds for a
broader class of systems. In particular, we conjecture a similar
identity for the XXZ spin chain\footnote{In this case the definition
  of normal ordering should include the lexicographic ordering of
  different $\widetilde{\mathbf{B}}$ and $\widetilde{\mathbf{A}}$
  operators.},
\begin{equation}
  \label{eq:18}
  \boxed{\hat{D}_{N,K}(v, q^{2v \partial_v}) =
    \mathopen: \det_{\mathrm{col}} \left(
      \mathbf{1} - \mathbf{T}_{\mathrm{XXZ}}(v) q^{2v \partial_v} \right) \mathclose:}
\end{equation}

The proof of the XXX/trigonometric Gaudin duality~\eqref{eq:12}
reduces to the classical calculation. We have already performed it in
the previous section not only for $q \to 1$ but for general $q$, i.e.\
for XXZ spin chains.

\paragraph{Trigonometric and reduced Gaudin models are equivalent.}
\label{sec:trig-vers-reduc}
In this section we diverge slightly to clarify the relationship with
our previous work on the subject. In~\cite{MMRZZ} we have proven the
duality between the XXX spin chain and the \emph{reduced} Gaudin
model. We now briefly describe why this model is equivalent to the
\emph{trigonometric} Gaudin model, at least classically. The quantum
case will not be discussed here, although we point out that the
quantization recipe obtained in~\cite{MMRZZ}\footnote{In that paper the
  universal difference operators (more precisely, the Baxter
  operators) were written in the Fourier dual variables. In our
  notation, the recipe consisted of putting all $x$ to the
  \emph{right}.} is formally conjugate to that of Eq.~\eqref{eq:13}.

The Lax matrix of the trigonometric Gaudin system is given by
\begin{equation}
   [L_{\mathrm{tG}}]_{ij}(w) = \delta_{ij} x_i + \delta_{i>j}
   \sum_{a=1}^K a_a^i b_a^j  +
    \sum_{a=1}^K \frac{w_a a_a^i b_a^j }{w - w_a} \,.
\end{equation}
It has linear Poisson bracket relations determined by the standard
classical trigonometric r-matrix. The reduced model is obtained from
the trigonometric one by means of a gauge transformation. The gauge
transformation matrix is given by
\begin{equation}
  g_{ij} = \delta_{ij} + \frac{\delta_{i > j}}{x_i - x_j}
  \sum_{a_i,\ldots , a_{j+1}} a_{a_i}^i \left( \delta_{a_i, a_{i-1}} + \frac{b_{a_i}^{i-1}
      a_{a_{i-1}}^{i-1}}{x_i - x_{i-1}} \right) \cdots \left(
    \delta_{a_{j+2} a_{j+1}} + \frac{b_{a_{j+2}}^{j+1}
      a_{a_{j+1}}^{j+1}}{x_i - x_{j+1}} \right)
  b_{a_{j+1}}^j\,,\notag
\end{equation}
\emph{The two systems are thus completely equivalent,
  $L_{\mathrm{tG}}(w) = gL_{\mathrm{redG}}g^{-1}$.} Note that the
Poisson brackets of the gauge transformed Lax matrix contain a
quadratic part. In the reduced Gaudin this part appeared from the
Dirac reduction.

\paragraph{Resum\'e and further perspectives.}
\label{sec:further-perspectives}

We conjecture the spectral duality for XXZ spin chains and prove it
for the classical case. We also obtain the universal difference
operator for the trigonometric Gaudin model and prove the quantum
XXX/Gaudin duality. We obtain a compact expression for
the universal difference operators using the notion of normal ordering
and conjecture the relationship between the classical and quantum
expressions. The equivalence of trigonometric and reduced Gaudin
models is also established.

Our proof of the duality is rather direct and does not rely on the
knowledge of the concrete (Bethe) eigenvectors of the
Hamiltonians. However, for the XXX and trigonometric Gaudin models
there is a beautiful theorem by Mukhin Tarasov and
Varchenko~\cite{MTV07} providing the correspondence between the kernel
of the universal difference operator and the Bethe vectors. One can
anticipate a similar correspondence for the XXZ spin chains.

Eventually, one hopes to fill the gap in the proof of the quantum
spectral duality for the XXZ spin chains by working out a proof of
the normal ordering identity~\eqref{eq:18}. It should go along the
same lines as that of Eqs.~\eqref{eq:14} and~\eqref{eq:17}.

From the point of view of the gauge theory the operators $\mathbf{A}$
and $\mathbf{B}$ correspond to certain operators supported on two
dimensional surfaces. The universal difference operator yields the
quantization conditions which determine the vacua of the worldvolume
theories of these surface operators. More concretely, the vacuum is a
state, for which the monodromy of the solution to the equation
$\hat{D}_{N,K} (v, q^{2v\partial_v}) |Q(v)\rangle = 0$ vanishes. From
the point of view of (the 3d extension of) the AGT
correspondence~\cite{3dAGT} surface operators are represented by
degenerate primary fields in the CFT living on the bare spectral
curve. One can, therefore, look for the implications of the spectral
duality for this CFT.

As an application to  integrable systems we also plan to study the
continuous limit of the spectral duality. In this limit we expect to
find relation between the continuous Heisenberg magnet (or more
generally 1+1 Hitchin systems \cite{LOZ}) and large $N$ limit of the
$\mathfrak{gl}_N$ Gaudin models proposed recently in \cite{AALOZ}.

It would be interesting to explore the relationship between spectral
duality and knot theory along the lines of the 3d/5d
duality~\cite{3d5d}. In this setting the spectral duality might be
connected with the skew Howe duality~\cite{Howe} recently used in
connection with link homologies in~\cite{Howe2}.

Let us also specifically mention the work~\cite{Bazhanov:2005as} in
which the same duality (called there the ``rank--size duality'')
emerged from the perspective of three dimensional integrable lattice
models. The connection between these models and five dimensional gauge
theories deserves further investigation.

\paragraph{Acknowledgements}
\label{sec:acnoledgements}

We would like to thank J.~Harnad, P.~Koroteev and A.~Varchenko for
helpful discussions. Our work was partly supported by
Ministry of Education and Science of the Russian Federation under
contract 8498, the Brazil National Counsel of Scientific and
Technological Development (A.Mor.), by NSh-3349.2012.2 (A.Mir.,
A.Mor.), by RFBR grants 14-02-92009, 13-02-00457 (A.Mir.), 13-02-00478
(A.Mor., Y.Z.), \mbox{12-02-31595} (Y.Z.) and 12-01-00482 (B.R.,
A.Z.), by joint grants 12-02-92108-Yaf (A.Mir., A.Mor.),
13-02-91371-ST (A.Mir., A.Mor., B.R.), 14-01-93004-Viet (A.Mir.,
A.Mor.), by leading young scientific groups RFBR 12-01-33071
mol$\_$a$\_$ved (B.R., Y.Z., A.Z.)  and by D.~Zimin's ``Dynasty''
foundation (A.Z.).

\small{

}
\end{document}